\documentclass[11pt]{article}

\newcommand{\be}{\begin{equation}}
\newcommand{\ee}{\end{equation}}
\newcommand{\kk}{$K\overline{K}$ }
\newcommand{\kpkm}{$K^+K^-$ }

\title{THEORETICAL MODEL OF THE $\phi$ MESON PHOTOPRODUCTION AMPLITUDES}
\author {L. LE\'SNIAK$^a$, A.P. SZCZEPANIAK$^b$   \\[4mm]
\small {$^a$Henryk Niewodnicza\'nski Institute of Nuclear Physics,}
 \\ \small{ PL 31-342 Krak\'ow, Poland} \\
 \small {$^b$Physics Department and Nuclear Theory Center,}
 \\ \small{ Indiana University, Bloomington, IN 47405, US } }
    
\begin{document}
\vspace{6mm} 

\maketitle
\vspace{6mm}
\begin{abstract}

P--wave amplitudes for elastic \kpkm photoproduction on hydrogen near the 
$\phi(1020)$ resonance have been derived in an analytical form and 
 a partial wave decomposition of the amplitudes has been performed. 
We discuss the high energy limit of the resulting amplitudes and compare 
 two types of pomeron coupling to nucleons. 
\end{abstract}

  
\vspace{6mm} 

\section{Introduction \label{introd}}    

Studies of photoproduction processes are very useful in application to
meson spectroscopy. The cross sections for vector 
meson photo- or electroproduction are relatively high since these processes
are diffractive at medium and high photon energies. Many aspects of 
photoproduction reactions of the $\rho$, $\omega$, $\phi$, $J/\Psi$ and other 
vector mesons have been studied
over a few decades \cite{Bauer} and these reactions continue attracting
 significant attention \cite{DIS}. After construction of the HERA
 collider, energies and momentum transfers accessible for photoproduction 
 have been substantially increased \cite{Abramowicz},\cite{ZEUS}.
 Here one should  underline many important contributions of Jan
 Kwieci\'nski in the field of high
energy electron and photon interactions (see, for example his contributions in 
\cite{DIS} and \cite{JK}).

Most of the known mesons are successfully classified as nonets of the 
$q\overline{q}$ quark model. The scalar mesons, however, escape 
 such a simple classification and are usually referred to 
 as "unusual" mesons. Their properties are
lively discussed in theoretical and experimental
papers (see, for example, a note on scalar mesons by S. Spanier and
N. T\"{o}rnqvist in \cite{PDG2002}). There is also a new proposed,  
  GlueX/Hall D  experiment at the energy upgraded Jefferson Laboratory
  in the US, to study production and decay characteristics of 
 gluonic excitations and unusual mesons \cite{CDR}. 

 Production of scalar mesons by photon beams is suppressed in comparison with
 production of vector mesons. In particular, these are the cases of the 
$f_0(980)$  or the $a_0(980)$ meson photoproduction. Nevertheless, even if
 the cross section for the $f_0(980)$ photoproduction was too small to
 be directly measurable, it is possible to see effects of the
 $f_0(980)$ through 
 interferences in the $K^+K^-$ decay channel of the 
 $\phi(1020)$ meson. Experiments exploring such effects have been done at DESY 
 \cite{Fries}, \cite{Behrend} and at the Daresbury Laboratory
 \cite{Barber}. In the first experiment, performed in the photon
 energy range from 4.6 to 6.7 GeV and at low momentum transfers squared 
 $|t| < 0.2\mbox{ GeV}/c^2$, the 
interference of $f_0(980)$ with $\phi(1020)$ has been observed
as a forward-backward asymmetry in the $K^+$ angular distribution. 
 This enabled to set the upper limit for the total $f_0(980)$
 photoproduction. In the Daresbury experiment, performed 
 at slightly lower energies, between 2.8 and 4.8 GeV and at much
 wider momentum transfers, $|t|$ up to 1.5 (GeV/c)$^2$, the
 interference between the dominant P--wave, corresponding to the
 $\phi(1020)$ production, and the S--wave has also been clearly seen. 
 The interpretation of the data, however was not unique. The authors 
 of ref. \cite{Barber} obtained values of the total $f_0(980)$
 photoproduction cross section between 10 and 100 nb, where the first 
 value was obtained under the  assumption  of a nonresonant S--wave 
 and the second,  for a resonating $f_0(980)$ contribution to the S--wave.

In 1998 the theoretical results of the coupled channel analysis of the S--wave 
$\pi\pi$ and \kk photoproduction at a few GeV laboratory photon
momentum have been  published \cite{foto}. The final state interactions in two
channels for the effective two-pion or two-kaon masses below and above the 
\kk threshold have been included. The momentum transfer distributions
 and the effective mass distributions as well as the total S-- wave
cross sections have been calculated. The results indicate 
 that the $\pi\pi$ and \kk photoproduction processes should be 
 experimentally accessible and 
 could provide  valuable information on the structure of the
 underlying scalar resonances.  
 Recently the experimental results on large momentum transfer 
 photoproduction of $\phi(1020)$ mesons using the CLAS detector
 at Jefferson Laboratory have been published \cite{Anciant} and
 low-momentum transfer data for $\pi^0\pi^0$ production are
 expected from the Hall B Radphi~\cite{Radphi} experiment. 

The component of the P--wave ($\phi(1020)$) amplitude which can
interfere with  the S--wave amplitude is different from the dominant P--wave
 amplitude. In most phenomenological models of $\phi$ 
 photoproduction only amplitudes  corresponding to the $\phi$ spin 
  projection equal to the photon spin projection,  +1 (or
  --1) were considered. This is a consequence of the s--channel 
 helicity conservation in peripheral production. The S--wave
  of the produced pair of kaons, however,  can interfere
 only with the non-dominant P--wave helicity amplitude which
 corresponds to the  final \kk pair angular momentum projection on the 
 quantization axis equal to  zero. Therefore, to treat the S--P interference
 phenomena of the $K^+K^-$ production in vicinity of the $\phi(1020)$ mass one
 has to make a model of all P--wave photoproduction amplitudes and not only of 
 the dominant helicity component. The  principal aim of this paper is
 to formulate a relevant model of these amplitudes (for other models
 of the $\phi$ meson  photoproduction, see for example  \cite{Soding}--
\cite{ TitovLee}, \cite{AT} and \cite{OB}. 

In chapter 2 we present a model of the elastic \kpkm photoproduction
on proton suitable for the photon energies well above the $\phi(1020)$ 
production threshold.
In chapter 3 we write explicit expressions for twelve independent spin 
amplitudes.  Chapter 4 is devoted to a discussion of the high energy
limit. In chapter 5  alternative pomeron exchange amplitudes are
compared to the amplitudes derived in chapter 3. Concluding remarks
are given in chapter 6.

\section{Model of the P--wave photoproduction of the \kpkm pairs \label{P--wave}}
    
Let us consider the elastic photoproduction of $K^+K^-$ pairs on the proton
from  the \kk threshold to the \kk effective mass range including 
 the $\phi(1020)$ resonance. We know from \cite{foto} that 
 8 amplitudes are needed to describe the S--wave of the $K^+K^-$
 system. In addition to the S--wave amplitudes we have to construct 24
P--wave amplitudes in the above effective mass range. 
They depend on two spin projections of the photon ($\lambda=\pm 1$), 
three spin projections of the $\phi$ ($ M = 1, 0 ,-1$), two incoming
proton spin projections ($s_1 = \pm 1/2$) and two final proton spin 
 projections  ($s_2 = \pm 1/2$). Parity conservation places
 further restrictions on the amplitudes and these will be discussed
 later. 
 If we include only S and P partial waves so that 
 the maximum value of the \kpkm angular momentum $J$ equals to one then the 
 reaction 
\be
\gamma + p \rightarrow K^+ + K^- + p          \label{reaction}
\ee
is described by 36 amplitudes $T^J_M(\lambda, s_1, s_2)$. We denote the 
photon four-momentum by $q$, the initial and
final proton four-momenta by $p_1$ and $p_2$ and the produced kaon $K^+$
and $K^-$ four-momenta by $k_1$ and $k_2$, respectively. The amplitudes   
$T^J_M(\lambda, s_1, s_2)$ depend on five independent kinematical variables
$s, t, M_{KK}$ and $\Omega_{K}$, where $s$ and $t$ are the Mandelstam variables
equal to square of the total energy in the center of mass system and the 
four-momentum 
transfer squared, respectively. $M_{KK}$ is the \kpkm effective mass and 
$\Omega_{K}$ denotes the $K^+$ solid angle. The \kk center of mass system, in
which ${\bf k_1}+ {\bf k_2}=0$, is a 
convenient frame of reference to discuss kaon angular distributions.  One 
can choose the z--axis antiparallel to the direction of the recoiling 
proton and  the y--axis parallel to ${\bf p_1 \times p_2}$, where 
 ${\bf p_1}$ is the initial proton momentum and ${\bf p_2}$ is the
 final proton momentum. Thus the photon 
 momentum ${\bf q}$ lies in the x--z plane being the $\phi$--proton 
 production plane. 
 We choose the x--axis in such a way that the x--component of ${\bf
   p_1}$ is positive and consequently the projection of ${\bf q}$ on
 the x--axis is negative.
This frame, called the s--channel helicity frame, is convenient to study the 
 wave interference. 

Let us assume that the main production mechanism of two kaons has  
a diffractive character and proceeds in two steps. In the first step the 
$\phi(1020)$ meson is produced on a proton and in the second step it
decays into two kaons. One can assume that the $\phi(1020)$ is produced via 
 soft pomeron exchange which leads to almost purely imaginary
 amplitude at small momentum transfers \cite{DL}. The \kk effective
 mass distribution of the $\phi$ decay can be simply related to the 
relativistic Breit- Wigner propagator
\be
BW(M_{KK})=1/(M_{\phi}^2-M_{KK}^2- i M_{\phi}\Gamma_{\phi}),      \label{BW}
\ee
where $M_{\phi}$ and $\Gamma_{\phi}$ are the $\phi(1020)$ meson mass and total
width.
We consider the following P--wave amplitude :
\be
T^P(\lambda, s_1, s_2)= F(s,t) BW(M_{KK}) \overline{u}(p_2,s_2)\gamma_{\alpha}
u(p_1,s_1)w_{\alpha}^{\lambda},                                      \label{TP}
\ee
where
\be
w_{\alpha}^{\lambda}=q_{\alpha}\epsilon^{\lambda}\cdot (k_1-k_2)-q\cdot(k_1-k_2)
\epsilon_{\alpha}^{\lambda}.
                                                       \label{w}
\ee
In (\ref{TP}) $F(s,t)$ is a phenomenological function which should be suitably
parameterized to reproduce the energy, $s$ and $t$-- dependence of the 
 experimental cross section 
for the $\phi(1020)$ photoproduction.  Its 
analytical form can depend on the $t$-- range covered in a given
experiment. The four--component Dirac spinors of incoming and
 outgoing protons are denoted by $u(p_1,s_1)$ and $u(p_2,s_2)$, respectively;
$\gamma_{\alpha}$, $\alpha=0,1,2,3$, are the Dirac matrices and 
$\epsilon^{\lambda}_{\alpha}$,
$\lambda=\pm 1$, are the photon polarization four-vectors.

Using the Dirac equations for proton spinors in the initial 
  and final states one can rewrite the P--wave amplitude in 
 an equivalent form,
\be
T^P(\lambda, s_1, s_2)= F(s,t) BW(M_{KK}) \overline{u}(p_2,s_2)\gamma_{\alpha}
u(p_1,s_1)v_{\alpha}^{\lambda},                                      \label{TPv}
\ee
where
\be
v_{\alpha}^{\lambda}=k_{\alpha}\epsilon^{\lambda}\cdot (k_1-k_2)-q\cdot(k_1-k_2)
\epsilon_{\alpha}^{\lambda}
                                                       \label{v}
\ee
and $k =k_1+k_2$ is the $\phi$ meson four-momentum.
The P--wave amplitude, introduced above, is gauge invariant. It can be expressed
as a sum of three partial waves $T^1_M(\lambda, s_1, s_2)$ corresponding to 
three different $\phi$ spin projections $M$, 
\be
T^P(\lambda, s_1, s_2)= \sum_{M=1,0,-1} T^1_M(\lambda, s_1, s_2) 
Y^1_M(\Omega_{K}),
                                                               \label{waves}
\ee
where $Y^1_M(\Omega_{K})$ are the spherical harmonics. 

 The resulting amplitudes are invariant under parity 
transformation. This property can be used to
eliminate 12 of the 24 helicity amplitudes since they satisfy the following
symmetry relations (valid also for the S--wave, $J=0$) \cite{SW}
\be
T^J_{-M}(-\lambda, -s_1, -s_2)=(-1)^{M-s_2-\lambda+s_1} 
~~T^J_M(\lambda, s_1, s_2).
                                                              \label{sym}
\ee
Using this symmetry one can consider only amplitudes with $\lambda=+1$
and then one can refer to  $T^1_1(1, s_1, s_2)$ as the non-flip amplitude,
$T^1_0(1, s_1, s_2)$ as the single-flip amplitude and 
$T^1_{-1}(1, s_1, s_2)$ as the double-flip amplitude. The first amplitude is
dominant and the second one is responsible for the S--P interference..

The four-fold differential cross section can be expressed in terms of the 
helicity amplitudes $T^J_M(\lambda, s_1, s_2)$ in the following way,
\be
\frac{d\sigma}{d\Omega dM_{KK} dt}= \frac{1}{4} \sum_{J,M,\lambda, s_1, s_2}
|T^J_M(\lambda, s_1, s_2) Y^J_M(\Omega_{K}|^2.                                   \label{dsig}
\ee

\section{Explicit form of the P--wave amplitudes \label{P--wave amplit.}}
    
Let us proceed to a derivation of the P--wave amplitudes. In the first 
 step one can evaluate the matrix element consisting of the 
Dirac matrices $\gamma_{\alpha}\equiv(\gamma_0,\mbox{\boldmath $\gamma$})$,
\be
\overline{u}(p_2,s_2)\gamma_{\alpha}u(p_1,s_1)v_{\alpha}=
 \overline{u}(p_2,s_2)(v_0\gamma_0-{\bf v} \cdot \mbox{\boldmath $\gamma$})
 u(p_1,s_1).
                                                              \label{vgamma}
\ee 
We get  
\be
\overline{u}(p_2,s_2) \gamma_0 u(p_1,s_1) =
f u^{\dagger}_{s_2} (A- i {\bf B} \cdot \mbox{\boldmath $\sigma$})u_{s_1},            \label{u0u}
\ee
\be
A=1+{\bf r_1 \cdot r_2},                             \label{A}
\ee 
\be
{\bf B}= {\bf r_1\times r_2},
                                                     \label{B}                                                      
\ee
\be
f=[(E_1+m)(E_2+m)]^{1/2}                           \label{f}
\ee
and
\be
\overline{u}(p_2,s_2){\bf v}^{\lambda} \cdot \mbox{\boldmath $\gamma$}
u(p_1,s_1)  = f ~~ u^{\dagger}_{s_2}(C^{\lambda} + {\bf D}^{\lambda} \cdot
 \mbox{\boldmath $\sigma$}) u_{s_1},
                                                        \label{uvgammau}
\ee 
\be                                                       
C^{\lambda} = {\bf v}^{\lambda} \cdot {\bf (r_1 + r_2)},  \label{Clambda}
\ee
\be
{\bf D}^{\lambda} = i \cdot [{\bf v}^{\lambda} \times ({\bf r_1 - r_2})].
                                                   \label{Dlambda}    
\ee

In (\ref{u0u}) and (\ref{uvgammau}) $u_{s_1}$ and $u_{s_2}$ are two-component 
Pauli spinors for the incoming and outgoing protons, 
and the Dirac spinors are normalized to $2m$, $m$ being the proton mass
; $E_1$ and $E_2$ are the energies of the incoming and 
outgoing proton, respectively. The vectors   
${\bf r_1}$ and ${\bf r_2}$ are defined as,
\be
{\bf r_1}=\frac{{\bf p_1}}{E_1+m},~~~~ {\bf r_2}=\frac{{\bf p_2}}{E_2+m}.
                                                                  \label{rr}
\ee
The three Pauli matrices are
denoted by $\mbox{\boldmath $\sigma$}$. The formulae (\ref{u0u}) till (\ref{rr})
are valid in any reference frame. 

In the following evaluation of the transition amplitudes we use the transverse 
gauge so the fourth component of the photon polarization 
$\epsilon_0^{\lambda}=0$.
In the $\phi$ center of mass frame one can express the four vector 
$v_\alpha^{\lambda}= (v_0^{\lambda},{\bf v}^{\lambda})$ as follows,
\be
v_0^{\lambda}= - 2 M_{KK}~\mbox{\boldmath $\epsilon$}^{\lambda} \cdot {\bf k_1},                
                                                                \label{v0}
\ee
\be
{\bf v}^{\lambda}=2 {\bf q \cdot k_1}~ \mbox{\boldmath $\epsilon$}^{\lambda}.     \label{vector}
\ee  
Let us denote the polar angle of the $K^+$ meson by $\theta$ and the azimuthal                                          
angle by $\phi$. The polar angle of the photon momentum ${\bf q}$ is called 
$\theta_q$, so the components of ${\bf q}$  are,  

\be
{\bf q}=|{\bf q}| (-\sin\theta_q, 0, \cos\theta_q).  \label{q}
\ee
 The photon polarization vectors are perpendicular to ${\bf q}$ and are equal to,
\be
\mbox{\boldmath $\epsilon$}^{\lambda}=-\frac{\lambda}{\sqrt{2}}(\cos\theta_q,
i\lambda,\sin\theta_q).
                                                      \label{epsilon}
\ee                                                      
Using the energy and momentum conservation one can calculate the kinematical 
variables expressed in the $\phi$ center of mass frame in terms of the
invariants $s$, $t$ and $M_{KK}$, 
\be
E_1=\frac{s-m^2+t}{2 M_{KK}},~~~~E_2=\frac{s-m^2-M_{KK}^2}{2 M_{KK}} \label{EE},
\ee
\be
|{\bf q}| = \frac{M_{KK}}{2}-\frac{t}{2 M_{KK}}        \label{|q|}
\ee 
and
\be
\cos\theta_q= \frac{E_1^2 - E_2^2 - |{\bf q}|^2}{2 |{\bf q}||{\bf p}_2|}.  
                                                                 \label{cosq}
\ee                                                     
The partial wave decomposition of the P--wave amplitude (\ref{TP}) is performed
by taking into account two identities valid for the scalar products present in 
(\ref{v0}) and in (\ref{vector}),
\be
\mbox{\boldmath $\epsilon$}^{\lambda} \cdot {\bf k_1}=\kappa (\frac{4\pi}{3})^{1/2}
\sum_{M=\pm1,0} b^{\lambda}_M ~Y^1_M(\Omega_{K}),              \label{epsk}
\ee
\be
b^{\lambda}_M = \delta^{\lambda}_M + \frac{\cos\theta_q - 1}{2}
(-1)^{\frac{\lambda-M}{2}} ~~~ {\rm for} ~M=\pm 1,                        \label{bpm1}
\ee
\be
b^{\lambda}_0=-\lambda \sin\theta_q/\sqrt{2},                       \label{b0}
\ee
and
\be
{\bf q} \cdot {\bf k}_1 = |{\bf q}|~ \kappa (\frac{4\pi}{3})^{1/2}
\sum_{M=\pm1,0} \rho_M ~Y^1_M(\Omega_{K}),       \label{qk1}
\ee
\be
\rho_1=\sin\theta_q/\sqrt{2},~~\rho_0=\cos\theta_q,
~~ \rho_{-1}=-\sin\theta_q/\sqrt{2}.                            \label{roM}
\ee
The quantity $\kappa$ in (\ref{epsk}) and (\ref{qk1}) is the magnitude
of the relative kaon momentum in the $\phi$ c.m. frame. This momentum
is related to the effective \kk mass,
\be
\kappa= (\frac{M_{KK}^2}{4}-m_K^2)^{1/2},                         \label{kappa}
\ee
where $m_K$ is the kaon mass.
The matrix elements (\ref{u0u}) and (\ref{uvgammau}) depend on the proton 
helicities $s_1$ and $s_2$, so they form $2 \times 2$ matrices, in which 
the first index $s_1=\pm1/2$ labels columns and the second one 
$s_1=\pm1/2$ labels rows.
Below we define four spin matrices and write their matrix elements,
\vspace{0.3cm}
 
\be 
\hspace{1cm} N_{s_1 s_2} \equiv u^{\dagger}_{s_2}u_{s_1}=
\vspace{-0.5cm}                                                 
  \left(  \begin{array}{cc}
              \tilde{s}     & \tilde{c}     \\
              -\tilde{c}    & \tilde{s}                       
          \end{array}           \right)~~,                 \label{N} \ee
          
\vspace{0.9cm}
 
\be 
\hspace{1cm} N_{s_1 s_2}^x  \equiv u^{\dagger}_{s_2}\sigma_x u_{s_1}=
\vspace{-0.5cm}                                                
  \left(  \begin{array}{cc}
              \tilde{c}     & -\tilde{s}     \\
              -\tilde{s}    & -\tilde{c}                       
          \end{array}           \right)~~,                   \label{Nx} \ee
          
\vspace{0.9cm} 

\be 
\hspace{1cm} N_{s_1 s_2}^y \equiv u^{\dagger}_{s_2}\sigma_y u_{s_1}=
\vspace{-0.5cm}                                                 
  \left(  \begin{array}{cc}
              i\tilde{c}     & -i\tilde{s}     \\
              i\tilde{s}     & i\tilde{c}                       
          \end{array}           \right)~~,  
                                                           \label{Ny} \ee
          
\vspace{0.9cm} 

\be 
\hspace{0.8cm} N_{s_1 s_2}^z  \equiv u^{\dagger}_{s_2}\sigma_z u_{s_1}=
\vspace{-0.5cm}                                                 
   \left(  \begin{array}{cc}
              -\tilde{s}     & -\tilde{c}     \\
              -\tilde{c}     & \tilde{s}                       
          \end{array}           \right)~~.                     \label{Nz} \ee
          
\vspace{0.9cm}          

We have introduced the following abbreviations in
 the matrices written above,
\be 
\tilde{c}=\cos({\theta_1/2}) , ~~~~  \tilde{s}=\sin({\theta_1/2}), \label{tilda}
\ee        
where $\theta_1$ denotes the polar angle of the initial proton in the s-channel
helicity frame. 
The following kinematical relation
together with formulae (\ref{EE}) can be useful in computations of 
 partial wave amplitudes,
\be
\cos\theta_1 = \frac{m^2-E_1~E_2-t/2}{|{\bf p}_1||{\bf p}_2|}.  \label{costeta1}
\ee
  The angle $\theta_1$ of the initial proton tends to 180$^o$ in the limit of
 high photon energy and at small momentum transfers. Let us remark here that the
 polar angle of the final proton is always equal to 180$^o$ in our reference
 frame.
 
Finally, putting together the expressions given by (\ref{TPv}--\ref{waves}),
(\ref{vgamma}--\ref{vector}), \mbox{(\ref{epsk}--\ref{roM})} and (\ref{N}--\ref{Nz}) one 
obtains the P--wave partial wave amplitudes,
\begin{eqnarray}
T^1_M(\lambda, s_1, s_2)= R~ [N_{s_1 s_2} (M_{KK} b^{\lambda}_M A +
|{\bf q}| \rho_M G^{\lambda} ) + N_{s_1 s_2}^x |{\bf q}| \rho_M 
{\bf H}^{\lambda}_x \nonumber \\
+N_{s_1 s_2}^y (- i M_{KK} b^{\lambda}_M| {\bf B}| + |{\bf q}| 
\rho_M {\bf H}^{\lambda}_y)+N_{s_1 s_2}^z |{\bf q}| \rho_M {\bf H}^{\lambda}_z],
                                                               \label{T1konc}
\end{eqnarray} 
where 
\be
G^{\lambda}= \mbox{\boldmath $\epsilon$}^{\lambda} \cdot ({\bf r}_1 + 
{\bf r}_2),
                                                               \label{G}
\ee 
$H^{\lambda}_x, H^{\lambda}_y, H^{\lambda}_z$ are the components of the
vector,
\be
\mbox{\boldmath $H$}^{\lambda} = i \mbox{\boldmath $\epsilon$}^{\lambda} \times 
({\bf r}_1 - {\bf r}_2)                                               \label{H}
\ee   
 and the common factors are included in the $R$ function defined as,
\be
R=-2 (\frac{4\pi}{3})^{1/2}\kappa f F(s,t) BW(M_{KK}).         \label{R}
\ee
\section{High energy limit \label{high}}    

It is instructive to calculate the photoproduction amplitudes in the
limit of high photon energy. This is a limit in which the invariant $s$
is much larger that the momentum transfer squared and it is also larger than
the square of the effective mass $M_{KK}$. Let us notice that in this limit the
energies or the momenta of the incoming and the outgoing protons in the $\phi$
c.m. frame
are of the same order.     Using the formulae (\ref{T1konc}--\ref{H}) and (\ref{B}) one can notice that
terms proportional to ${\bf B}$, $D^\lambda_x$, $D^\lambda_y$ and
$D^\lambda_z$ vanish as $1/s$ in the limit  $s \to \infty$.
 It is sufficient
 to discuss only one
 case of the photon polarization $\lambda=1$. Due to the symmetry property 
(\ref{sym}) the amplitudes with  
$\lambda=-1$ are related to the previous case of $\lambda=1$ . Next one can 
verify that the coefficients $b^1_{\lambda}$ behave as follows,
\be
b^1_1 \approx 1,~~~b^1_0 \approx -\sqrt{-2 t}/M_{KK},
~~~ b^1_{-1} \approx -t/M_{KK}^2,                            \label{wspb}
\ee
since
\be
\cos{\theta_q}\approx \frac{1+t/M_{KK}^2}{1-t/M_{KK}^2}.   \label{costetaq}
\ee
The presence of the symbol $\delta^{\lambda}_M$ in (\ref{bpm1}) means that 
the $\phi$ 
amplitude, which conserves the helicity of the photon, dominates at high energy
and the low momentum transfers (the so-called s-- channel helicity conservation). 
Next we see that in the same limit,
\be
T^1_1(1,s_1,s_2)=2 R N_{s_1 s_2} M_{KK},               \label{T11}
\ee
\be
T^1_0(1,s_1,s_2)=-R N_{s_1 s_2} \sqrt{-2 t}             \label{T10}
\ee
and
\be
T^1_{-1}(1,s_1,s_2)\approx 0(1/s).         \label{T1-1}                
\ee
From (\ref{T11}) and (\ref{T10}) one notices that the $\phi$ production with 
helicity 0 is reduced if $-t$ is much
smaller than the square of the effective mass $M_{KK}$. The
amplitude $T^1_{-1}$ is of the order of $1/s$ and is negligible at high 
energies. 

Let us make a comment on the behavior of production amplitudes as functions
 of the proton helicities. In the high energy limit one has,
\be
\tilde{s} \approx 1, ~~~~\tilde{c} \approx  M_{KK}\sqrt{-t}/s . \label{cs}
\ee
Inspecting the expression in (\ref{N}) for $N_{s_1 s_2}$  we
 immediately notice dominance of the proton helicity non-flip 
 amplitudes in the \kk photoproduction
 over proton helicity flip amplitudes if $s \gg M_{KK}\sqrt{-t}$. 
\section{"Scalar" pomeron versus "vector" pomeron  \label{scalar pom.}}
   
In S\"{o}ding's paper \cite{Soding} one can find an expression for the 
diffractive $\rho$ 
photoproduction amplitude on a nucleon which is different from the P~--wave
amplitude written in (\ref{TP}) or in (\ref{TPv}).
 While the later 
amplitude has a typical "vector" or photon-like coupling of the exchanged 
pomeron to the nucleon the former amplitude is typical for the "scalar" pomeron
coupling. Such "scalar" type amplitudes have been used in description
 of the $\phi$ photoproduction in \cite{Williams} and in \cite{TitovLee}. The
 S\"{o}ding type amplitude adapted for the $\phi$ photoproduction is 
proportional to 
\be
U(\lambda, s_1, s_2)= V_{\lambda}~\overline{u}(p_2,s_2)u(p_1,s_1) ,
                                                                  \label{ampl.S}
\ee
where
\be
V_{\lambda}=q \cdot k ~\epsilon^{\lambda}\cdot (k_1-k_2)-q\cdot(k_1-k_2)~
\epsilon^{\lambda} \cdot k.
                                                       \label{V}
\ee
This amplitude has a factorized form.
The factor $V_{\lambda}$ in the $\phi$ c.m. frame is given by,
\be
V_{\lambda}= - 2 M_{KK}~|{\bf q}|~ \mbox{\boldmath $\epsilon^{\lambda}$}
 \cdot {\bf k_1}.
                                                                      \label{Vl}
\ee                                                                      
 The nucleon part which depends only on the proton spin projections reads,
\be 
\overline{u}(p_2,s_2)u(p_1,s_1)=f[(2-A)N_{s_1 s_2} +i |{\bf B}|N_{s_1 s_2}^y].
                                                               \label{uu}
\ee 
We make the partial wave decomposition of $U(\lambda, s_1, s_2)$
as follows,
\be
U(\lambda, s_1, s_2)=  \sum_{M=1,0,-1} U^1_M(\lambda, s_1, s_2) 
Y^1_M(\Omega_{K}).
                                                               \label{Uwaves}
\ee
Then the partial wave amplitudes in the "scalar" pomeron model read, 
\be
U^1_M(\lambda, s_1, s_2)= Z~b^{\lambda}_M [(2-A)N_{s_1 s_2} +i |{\bf B}|
N_{s_1 s_2}^y],     \label{U1M}                                                        
\ee                                                               
where
\be
Z=-2 (\frac{4\pi}{3})^{1/2}\kappa f M_{KK} ~|{\bf q}|.     \label{zet}
\ee
The calculation of the high energy limit specified in the previous chapter can
be done using the explicit form of the functions $A$ and ${\bf B}$ defined
in (\ref{A}), (\ref{B}), (\ref{rr}) and (\ref{wspb}). One obtains, 
\be
U^1_M(\lambda, s_1, s_2)= Z~b^{\lambda}_M \frac{4mM_{KK}}{s} 
[N_{s_1 s_2} +i \frac{\sqrt{-t}}{2m}N_{s_1 s_2}^y].
                                                         \label{U1Mwys.en.}
\ee
We notice  that the amplitudes corresponding to the "scalar" pomeron exchange
are suppressed as $1/s$ at high energies. There is also another difference
between (\ref{U1Mwys.en.}) and (\ref{T10}--\ref{T1-1}), namely the proton
helicity flip contributions coming from the second part of (\ref{U1Mwys.en.})
proportional to $N_{s_1 s_2}^y$, can be important at $-t$ comparable to $m^2$.  

\section{Concluding remarks \label{conclusions}}
   
 Photoproduction of kaon pairs with
 invariant mass near $1\mbox{ GeV}$ is dominated by the $\phi(1020)$ resonance.
 This is close to the $K{\bar K}$ threshold and through S-- and P-- wave 
 interference can give important insight into the soft meson-meson 
 interactions and a possibility of formation of mesonic bound states. 
 The entangled P--wave $K{\bar K}$ state from the $\phi(1020)$ decay 
  has been recognized as a tool for  studies of $CP$ and possible
  $CPT$  violations~\cite{CP}, and the possibility of using 
  both S-- and P-- wave combinations opens up more
  possibilities for such studies~\cite{CPT}. 
  
  As the first step we have 
 derived analytical
  expressions for the elastic amplitudes for
 $\phi$ photoproduction. The main
 result is given by eq. (\ref{T1konc}). From the above formula one can 
 calculate 24 P--wave amplitudes depending on the helicities of the
photon, the \kk pair,
and the incoming and outgoing protons. These amplitudes can be used in the 
partial wave analysis of the \kpkm photoproduction for the \kpkm effective
masses varying from the \kk threshold to the range covering the $\phi(1020)$
resonance. In particular one can study the S--P interference effects if the 
S--wave amplitudes from ref. \cite{foto} are included. This will be a 
subject of the subsequent investigations. 
\section{Acknowledgments} 

  One of us (L. L.) would like to thank Jan Kwieci\'nski for several discussions
about the high-energy photoproduction processes.
 We thank Chueng - Ryong Ji and Robert Kami\'nski for their collaboration on the
 S--wave photoproduction and for an exchange of ideas on the P--wave amplitudes
in the early stage of this work.

 This work was supported in part by the US Department of Energy under
contract DE-FG02-87ER40365. 


\end{document}